\documentclass[11pt,a4paper]{article}
\pdfoutput=1
\usepackage{jheppub}

\usepackage{graphics, setspace}
\usepackage{slashed}

\usepackage{color}
\usepackage{graphicx}                   
\usepackage{bbm}
\usepackage{amsmath}
\usepackage{amssymb}
\usepackage{bbold}
\usepackage{xspace}
\usepackage{verbatim}
\usepackage{bm}
\usepackage{dcolumn}
\usepackage{tabu} 

\newcommand{\zcut}{z_{\text{cut}}}

\newcommand{\mathsym}[1]{{}}
\newcommand{\unicode}[1]{{}}
\newcommand{\be}{\begin{equation}}
\newcommand{\ee}{\end{equation}}
\newcommand{\bea}{\begin{align}}
\newcommand{\eea}{\end{align}}

\newcommand{\lb}{\Big{\lbrack}}
\newcommand{\rb}{\Big{\rbrack}}
\newcommand{\lp}{\Big{(}}
\newcommand{\rp}{\Big{)}}
\newcommand{\lbc}{\Big{\lbrace}}
\newcommand{\rbc}{\Big{\rbrace}}

\newcommand{\bl}[1]{\boldsymbol{\mathrm{#1}}}
\newcommand{\nn}{\nonumber}

\newcommand{\bmax}{b_{\text{max}}}
\newcommand{\cG}{\mathcal{G}} 

\begin{document}

\preprint{LA-UR-17-31338}

\title{Probing Transverse-Momentum Dependent Evolution With Groomed Jets}

\author{Yiannis Makris,}%
\author{Duff Neill,}
 \author{and Varun Vaidya,}%
\affiliation{%
Theoretical Division, MS-248, Los Alamos National Laboratory, Los Alamos, NM 87545
}%

\emailAdd{yiannis@lanl.gov}
\emailAdd{duff.neill@gmail.com}
\emailAdd{vvaidya@lanl.gov}

\abstract{
We propose an observable which involves measuring the properties (transverse momentum $p_{h\perp}$ and energy fraction $z_h$) of an identified hadron inside a groomed jet. The jet is identified with an anti-kT/CA algorithm and is groomed by implementing the modified mass drop procedure with an energy cut-off parameter, $z_{\text{cut}}$. The transverse momentum of the hadron inside the jet is measured with respect to the groomed jet axis. We obtain a factorization theorem in the framework of Soft Collinear Effective Theory (SCET), to define a Transverse Momentum Dependent Fragmenting Jet Function (TMDFJF). The TMDFJF is factorized into collinear and collinear soft modes by matching onto SCET$_+$. We resum large logarithms in $E_J/p_{h\perp}$, where $E_J$ is the ungroomed jet energy, to NLL accuracy and apply this formalism for computing the shape of the $p_{h\perp}$ distribution of a pion produced in an $e^+ +e^-$ collision. We observe that the introduction of grooming makes this observable insensitive to non-global logarithms and particularly sensitive to non-perturbative physics of the transverse momentum dependent evolution at low values of $p_{h\perp}$, which can be probed in the variation of the cut-off parameter, $z_{\text{cut}}$, of the groomer. We discuss how this observable can be used to distinguish between non-perturbative models that describe universal TMD evolution and provide a window into the three dimensional structure of hadrons.}

\maketitle


\section{Introduction}
The transverse momentum spectrum with respect to a fiducial axis of an energetic or massive color-singlet state has been recognized as an observable of fundamental interest in probing quantum chromodynamics (QCD) and the factorization of infrared effects from ultraviolet hard processes. In  hadron-hadron or hadron-lepton collisions (that is, Drell-Yan like spectra and semi-inclusive production of a hadron in deep inelastic scattering), when the fiducial axis is taken to be the momentum of the hadronic beam(s), these observables give a three-dimensional picture of the single parton dynamics within the hadronic object. The infrared dynamics can be factored into distribution functions sensitive to the produced transverse momentum and the energy deposited into the creation of the hard color singlet state\cite{Collins:2011zzd,GarciaEchevarria:2011rb,Chiu:2012ir} with appropriate soft factors and subtractions \cite{Manohar:2006nz}. This extends the traditional factorization of hadronic structure in terms of collinear Parton Distribution Functions (PDFs)\cite{Georgi:1977mg,Ellis:1978ty,Collins:1981uw,Collins:1989gx}. When the color-singlet state is within a final state jet not aligned with the beam (for example, observing the transverse momentum spectrum of a final state hadron within an $e^+e^-$-collision using hemispherical jets \cite{Collins:1981uk}), we can likewise extend the notion of collinear Fragmentation Functions (FFs) to include the relative motion of the hadron with respect to all of the other jet constituents. A critical feature of all of these transverse momentum spectra are their sensitivity to soft processes, where one can show that the equations governing the resummation of large logarithmic corrections assumes a universal form despite the very different scattering processes (Drell-Yan production, semi-inclusive deep inelastic scattering, and $e^+e^-$ to hadrons). This resummation of the generalized transverse-momentum dependent parton distribution (TMDPDFs) or transverse-momentum dependent fragmentation functions (TMDFFs) is often termed the transverse-momentum dependent evolution (TMD-evolution) \cite{Aybat:2011zv}. The goal is to predict how the spectrum changes with the energy scale of the underlying hard process that creates the massive or energetic color-singlet state. Within perturbation theory, the equations governing this evolution has reached a very precise determination \cite{Li:2016axz,Li:2016ctv,Vladimirov:2016dll}, allowing one to confidently investigate the non-perturbative dynamics of the recoiling radiation. 

However, though formally the TMD-evolution of a TMDFF function is identical to that of a TMDPDF, at a hadron-hadron collider, such fragmentation processes are often studied within a \emph{jet}.\footnote{See Refs.~\cite{Procura:2009vm,Jain:2011xz,Jain:2012uq,Bauer:2013bza,Dasgupta:2014yra,Baumgart:2014upa,Kaufmann:2015hma,Kang:2016ehg,Kang:2016mcy,Chien:2015ctp,Bain:2016rrv,Dai:2016hzf,Dai:2017dpc,Kang:2017glf,Kang:2017btw,Kang:2017mda} for recent work on fragmentation processes both generating and within jets.} Since jets are not an intrinsic object to QCD, but rather a pattern of radiation that is most likely to occur, one must both theoretically and experimentally use a precise, though ultimately arbitrary, jet definition. While all reasonable jet definitions can be shown to group together the same energetic radiation into the jet region, one necessarily selects for different configurations of soft radiation that will be included in the jet, and this can spoil the equivalence of the TMD-evolution between final state (TMD-FF's) and initial state processes (TMD-PDF's). Generically, one worries about soft correlations that span the whole event, entangling the pattern of soft radiation within the jet to either the underlying event with multiple parton interactions within the colliding hadrons~\cite{Forshaw:2012bi,Catani:2011st,Rogers:2010dm,Zeng:2015iba,Gaunt:2014ska,Rothstein:2016bsq}, or non-global color correlations arising from out-of-jet radiation radiating back into the measured jet, all of which are color connected back to the hard process~\cite{Banfi:2002hw,Dasgupta:2001sh}. Indeed, both effects could potentially spoil the factorization predictions for TMD-spectra found in~\cite{Kang:2017mda,Kang:2017glf}. Thus naively, one suspects that only the TMD-evolution of fragmentation processes within hemisphere jets at an $e^+e^-$ machine could be tied to the TMD-evolution of the Z-boson spectrum. 

Developments in jet substructure\footnote{See Ref.~\cite{Larkoski:2017jix} for a comprehensive review.} have shown that the modified mass drop tagging algorithm (mMDT) or soft-drop grooming procedure robustly removes contamination from both underlying event and non-global color-correlations, see Refs.~\cite{Larkoski:2014wba,Dasgupta:2013via,Dasgupta:2013ihk}, and have been applied to study a wide variety of QCD phenomenology within jets \cite{Larkoski:2017iuy,Hoang:2017kmk,Marzani:2017mva,Dasgupta:2016ktv,Frye:2016okc,Frye:2016aiz,Larkoski:2015lea,Dasgupta:2015lxh,Chien:2014nsa}. Exploiting this fact, we will give a concrete proposal as to how one can observe the universal TMD-evolution within these groomed jets, where we specify that we study the transverse momentum spectrum of a hadron within the jet with respect to the total momentum of the groomed jet, that is, all particles that pass the mMDT or soft-drop procedure.

The outline of the paper is as follows: we briefly review the mMDT/soft-drop procedure, then we follow the factorization arguments of Ref. \cite{Frye:2016okc,Frye:2016aiz} and give the factorization theorem for the groomed TMD-spectrum of a fragmented hadron, as well as related jet shape observables. We present the structure of the anomalous dimensions for the various objects found in the factorization theorem, working in the framework derived in Ref. \cite{Chiu:2012ir}, which governs the TMD-evolution. We then show how the normalized and groomed TMD-spectrum gives direct access to the rapidity renormalization group/Collins-Soper evolution, as a function of the grooming parameter $\zcut$.\footnote{For a discussion of the connection between the rapidity renormalization group and the Collins-Soper equation, see \cite{Collins:2017oxh}}

\section{Modified Mass Drop}
The modified mass-drop procedure \cite{Dasgupta:2013via,Dasgupta:2013ihk} or its generalization known as soft-drop \cite{Larkoski:2014wba} removes contaminating soft radiation from the jet by constructing an angular ordered tree of the jet through the Cambridge/Aachen (C/A) clustering algorithm \cite{Ellis:1993tq,Catani:1993hr,Dokshitzer:1997in,Wobisch:1998wt,Wobisch:2000dk}, and removing the branches at the widest angles which fail an energy requirement. As soon as a branch is found that passes, this branch is declared the groomed jet, and all constituents of the branch are the groomed constituents. What is remarkable about the procedure, is that it gives a jet with essentially zero angular area, since at large angles, all collinear energetic radiation is to be found at the center of the jet, and no cone is actually imposed to enclose this core. One simply finds the branch whose daughters are sufficiently energetic. Formally the daughters could have any opening angle, though their most likely configuration is collinear.

The strict definition of the algorithm is as follows. Given a ungroomed jet, first we build the clustering history: we start with a list of particles in the jet. At each stage we merge the two particles within the list that are closest in angle\footnote{This merging is usually taken to be summing the momenta of the particles, though one could use winner-take-all schemes \cite{Salam:WTAUnpublished,Bertolini:2013iqa,Larkoski:2014uqa}.}. This gives a pseudo-particle, and we remove the two daughters from the current list of particles, replacing them with the merged pseudo-particle. This is repeated until all particles are merged into a single parent. Then we open the tree back up. At each stage of the declustering, we have two branches available, label them $i$ and $j$. We require:
\begin{align}\label{SD:condition}
\frac{\text{min}\{E_i,E_j\}}{E_i+E_j}>\zcut\,,
\end{align}
where $\zcut$ is the modified mass drop parameter, and $E_i$ is the energy of the branch $i$. If the two branches fail this requirement, the softer branch is removed from the jet, and we decluster the harder branch, once again testing Eq. \eqref{SD:condition} within the hard branch. The pruning continues until we have a branch that when declustered passes the condition \eqref{SD:condition}. All particles contained within this branch, whose daughters are sufficiently energetic, constitute the groomed jet. Intuitively we have identified the first genuine collinear splitting.

For a hadron-hadron collision, one uses the transverse momentum with respect to the beam for the condition of Eq. \eqref{SD:condition}:
\begin{align}\label{SD:condition_pp}
\frac{\text{min}\{p_{Ti},p_{Tj}\}}{p_{Ti}+p_{Tj}}>\zcut\,.
\end{align}

We formally adopt the power counting $\zcut \ll 1$, though typically one chooses $\zcut\sim 0.1$. See \cite{Marzani:2017mva} for a study on the magnitude of the power corrections with respect to $\zcut$ for jet mass distributions.
\section{Factorization With Grooming}

The overall impact of the mMDT grooming is that we force ourselves into a regime that is dominated by purely collinear physics. Thus the properties of the jet\footnote{Formally, we mean the jet function appearing in Eq. \eqref{eq:SD_factorization}.} can be considered in isolation from the rest of the event in which the jet occurs, and depend only upon the flavor of the initiating parton, a fact that typically is only true for the energy spectrum of hadrons or jets. This is to say, that while the jet is of course color-connected to the rest of the event, the color charge and flavor of the initiating hard parton dominate the spectrum of groomed observable. In terms of factorization, the rest of the event appears as a single wilson pointed in the anti-collinear direction of the jet. Genuine soft color correlations from multiple jets at wide angles are a power suppressed contribution to the groomed spectrum, due to the grooming procedure putting one in a collinear factorization regime, see Ref. \cite{Frye:2016aiz}. This is exactly analogous to the fragmentation spectrum at moderate energy fractions of the fragmented hadron, which is also set by the color charge and flavor of the parton initiating the fragmentation (as encoded by the fragmentation function), and the complicated multi-jet soft correlations are power-suppressed. This allows us to write the following factorization for a mMDT groomed jet:
\small
\begin{align}\label{eq:SD_factorization}
\frac{d\sigma}{d^3\vec{p}_Jd\mathcal{M}}\Big(\zcut,R,C\Big)&=F_g\Big(Q,R,\zcut,\vec{p}_J,C\Big)J_{g}\Big(\mathcal{M},\zcut,R,E_J\Big)\nn\\
&+\sum_{q}F_q\Big(Q,R,\zcut,\vec{p}_J,C\Big)J_{q}\Big(\mathcal{M},\zcut, R, E_J\Big)+...\,.
\end{align}
The functions $F_g$ and $F_{q}$ represent the gluon and quark fractions of the scattering process initiating the jet to be studied.\footnote{The groomed spectra are set by the jet functions which by collinear factorization depend only the color charge of the initiating parton, however, the number of initiating quarks and gluons are still sensitive to the soft correlations in the scattering process, conditioned on the cuts defining the jet. These flavor fractions therefore cannot in general be considered in isolation of the rest of the event.} These fractions are dependent upon the center of mass energy, $Q$, of the collision, the jet momentum and energy, $\vec{p}_J$, and $E_J$,\footnote{In hadron-hadron collisions, this is equivalent to the rapidity, azimuth, and transverse momentum with respect to the beam of a centrally located jet.} the jet radius $R$, and the grooming parameter, $\zcut$. The parameter $C$ represents any other cuts or constraints one makes on the scattering process outside the groomed jet. The underlying hard scattering process can be either an exclusive or inclusive jet cross-section, with various and complicated additional vetoes or observed decay channels imposed or not. 

The parameter(s) $\mathcal{M}$ represents on the other hand all of the substructure measurements to be performed upon the groomed jet. Since this is the interesting part of the cross section, henceforth, we will omit writing the differential $d^3 \vec{p}_J$ explicitly. The jet functions $J_q$ and $J_g$ will be given an operator definition below, and at this stage in the factorization may contain large logarithms, which would be resummed using an additional factorization within the jet function itself. For our purposes we wish to understand the factorization and resummation structure of spectrum of hadron production within the groomed jet. Specifically, we will consider the energy fraction spectrum, $z_h$, of the hadron and its transverse momentum with respect to the direction of the total momentum of groomed jet\footnote{This is a distinct observable from the transverse momentum of the hadron with respect to a soft-insensitive axis like the winner-take-all axis \cite{Neill:2016vbi}. Both observables enjoy a form of collinear factorization and hence universality, though the spectra and resummation structure are distinct.}. 

\begin{figure}[h!]
 \centerline{\includegraphics[width = 0.7 \textwidth]{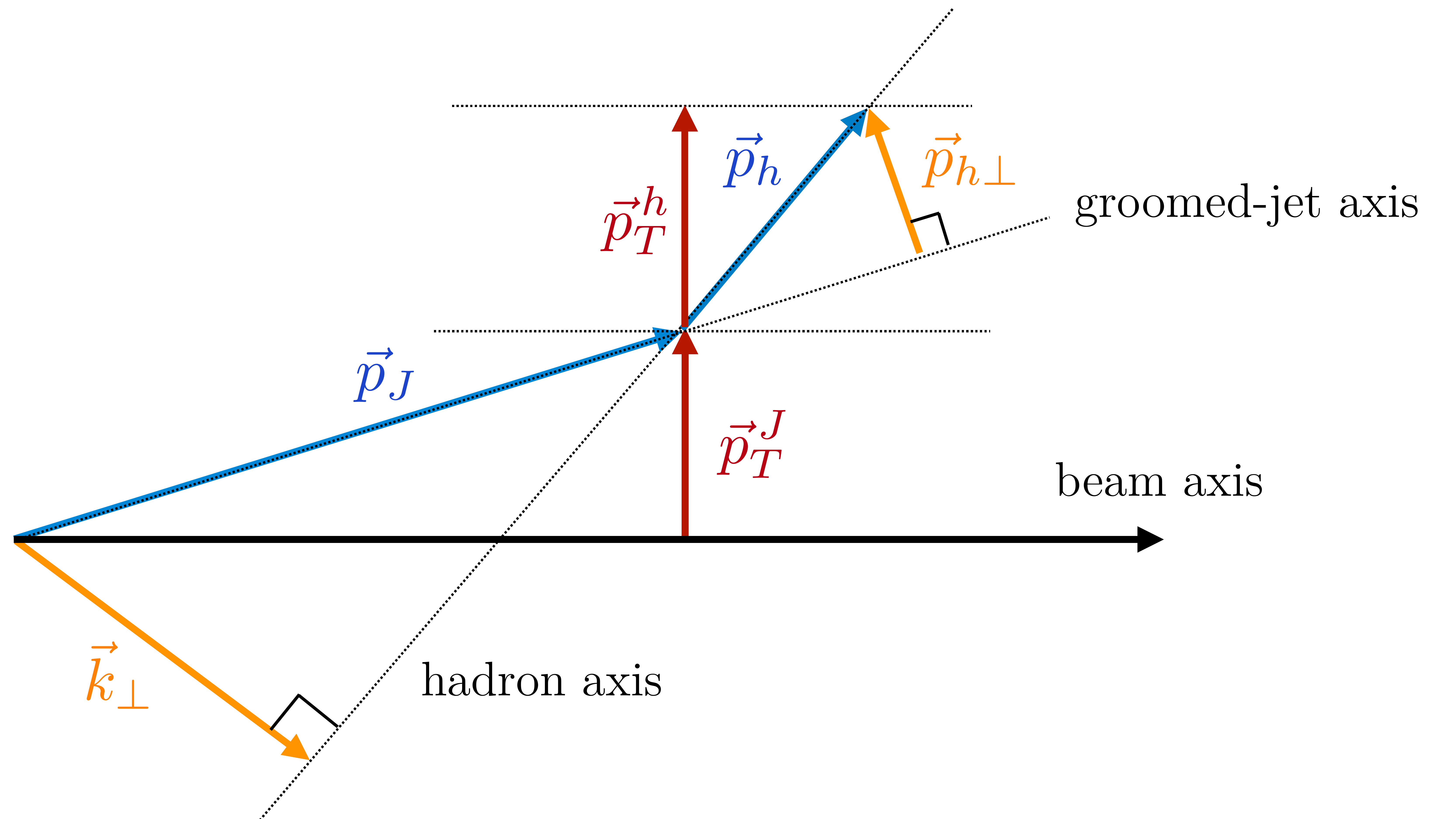}}
\caption{The geometric configuration of the jet and hadron axis relative  to beam. Here the jet axis is defined as the momentum of all the particles clustered by the jet algorithm. The vectors $\vec{k}_{\perp}$ and $\vec{p}_{h \perp}$ are two dimensional arrays with components as measured from the corresponding axis. }
\label{fig:diagram}
\end{figure}

Let us assume that the parton initiating the measured groomed jet is a quark (the argument that follows will apply equally well for any other parton flavor). We measure the transverse momentum of an identified hadron ($p_{h\perp}$) inside the jet carrying a fraction $z_h$ of the ungroomed jet energy.\footnote{The energy difference between the total energy of the groomed jet constituents and the energy of the ungrooomed jet is a collinear unsafe observable \cite{Larkoski:2015lea}, however, the additional constraint of the measured transverse momentum of the hadron provides a physical collinear cutoff.} We can then write down the jet function in the factorization of Eq. \eqref{eq:SD_factorization}: 
\begin{align}\label{eq:unfactorized_SD_jet_function}
&\cG_{q/h}\Big(z_h,\vec{k}_{\perp},\zcut,R,E_J\Big) \nonumber\\
&\qquad = z_h\sum_{X\in \text{Jet}(R)} \frac{1}{2N_c} \delta( 2E_J - p_{X}^{-}-p_h^-) \text{tr}\left[\frac{\slashed {\bar n}}{2} \langle 0| \delta^{(2)}(\vec{k}_{\perp} + \vec{\mathcal{P}}_{\perp}^{SD}) \chi_n (0)|X h\rangle \langle X h|\bar \chi_n(0)|0\rangle \right] \,.
\end{align}
where $\vec{k}_{\perp}$ is the jet transverse momentum with respect to the direction of motion of the identified hadron, $h$.
Although eventually we are interested in the hadron's transverse momentum with respect to the jet axis, $\vec{p}_{h \perp}$, we choose to work with $\vec{k}_{\perp}$ since this significantly simplifies our analysis. In this approach one needs to project the jet traverse momentum on the hadronic axis without need to worry about the recoiling hadrons from the fragmentation process.   In the collinear limit ($\vert \vec{p}_{h \perp} \vert/ E_J \ll 1$) we can relate $\vec{k}_{\perp}$ and  $\vec{p}_{h \perp}$ using a simple geometric argument. Using $\vert \vec{p}_{h} \vert \simeq z_h E_J $ we then have  $\vec{k}_{\perp} = \vec{p}_{h\perp}/z_h$ (see Fgure~\ref{fig:diagram}). 
Here $X$ contains all the particles in the jet of radius $R$ (assuming an appropriate jet definition, denoted Jet$(R)$, like the anti-$k_t$ algorithm\cite{Cacciari:2008gp}). The components of $\vec{k}_{\perp}$ are set by the label momentum operator $\vec{\mathcal{P}}_{\perp}^{SD}$ which projects the traverse momentum of the subset of those particles which pass the mMDT/soft-drop grooming procedure of the state $|X,h\rangle$. The sum over X includes integrals over the phase space of $X$. Note that the transverse momentum is defined as the recoil against all the particles which pass the grooming requirement. This implicitly defines the groomed jet axis to be the axis such that the sum of all transverse momenta with respect to that axis is zero.

This jet function contains large logarithmic corrections associated with the scales $|\vec{p}_{h\perp}|$, and $E_J$ when we are in the hierarchy:
\begin{align}
|\vec{p}_{h\perp}| \ll E_J\,.
\end{align}
Moreover, we must assume that the fragmented hadron is within the set of particles which pass the grooming, which further requires:
\begin{align}
\lambda^2=\Big(\frac{|\vec{p}_{h\perp}|}{E_J}\Big)^2\ll \zcut,\qquad R\sim O(1)\,.
\end{align}
We introduce light-cone jet direction $n=(1,\hat{n}_J)$, where $\hat{n}_J$ is the direction of the total momentum of the particles which pass mMDT/soft-drop. We also have $\bar{n} =(1,-\hat{n}_J) $, which is the conjugate direction, and finally, the transverse momentum plane to these directions. Any momentum $p$ can be decomposed in terms of these directions as
\begin{align}\label{eq:power_counting}
& p =( \bar{n}\cdot p, n\cdot p, p_{\perp})\,
\end{align}
The region of phase space which contributes to our measurement of $p_{h\perp}$, naturally gives two distinct power counting regions contributing to the observable,
\begin{align}
\text{collinear}: &\, p_c\sim E_J(1,\lambda^2,\lambda)\nonumber\\
\text{collinear-soft}: &\, p_{cs}\sim E_J\zcut\Big(1,\frac{\lambda^2}{\zcut^2},\frac{\lambda}{\zcut}\Big)
\end{align}
Within this region of phase-space, by following the logic the collinear/collinear-soft factorization arguments of so-called SCET$_+$ \cite{Bauer:2011uc,Larkoski:2015zka,Pietrulewicz:2016nwo}, we may further factorize the jet function in Eq.\eqref{eq:unfactorized_SD_jet_function} as (leaving the flavor generic):
\begin{align}
\label{Jet}
\cG_{i/h}\Big(z_h,\vec{k}_{\perp},\zcut,R,E_J\Big)&= z_h\int d^2 \vec{k}_{c \perp} \int d^2 \vec{k}_{s \perp} \delta^2 \left( \vec{k}_{\perp}+ \vec{k}_{c \perp}+ \vec{k}_{s\perp}\right)\tilde{\mathcal{D}}^{\perp}_{i/h} \left(z_h, E_J, \vec{k}_{c \perp}\right)\nn\\
&\times S_i^{\perp}\left( \vec{k}_{s\perp}, z_{cut}\right) + \mathcal{O} \lp\frac{\zcut}{R} \rp
\end{align}
The operator definitions of these functions are:
\begin{align}
\label{collinear}
\mathcal{D}_{q/h}^{\perp}(z_h,\vec{k}_{c\perp},E_J)&= \sum_{X} \frac{z_h}{2N_c} \delta( 2E_J - p_{Xh}^{-}) \text{tr}\lb\frac{\slashed {\bar n}}{2} \langle 0|\delta^{(2)}(\vec{k}_{c\perp}-\vec{\mathcal{P}}_{\perp}) \chi_n (0)|X h\rangle \langle X h|\bar \chi_n(0)|0\rangle \rb_{\vec{p}_{h \perp} = 0} \,,\\
\mathcal{D}_{g/h}^{\perp}(z_h,\vec{k}_{c\perp},E_J)&= \sum_{X} \frac{z_h}{2(N_c^2-1)} \delta( 2E_J - p_{Xh}^{-}) \text{tr}\lb\langle 0|\delta^{(2)}(\vec{k}_{c\perp}-\vec{\mathcal{P}}_{\perp}) B_{n\perp}^{\mu} (0)|X h\rangle \langle X h|B_{n\perp\mu}(0)|0\rangle \rb_{\vec{p}_{h \perp} = 0} \,,\\
S_{i}^{\perp}(\vec{k}_{s\perp}, E_J,\zcut)&=\frac{1}{N_i}\text{tr} \lb \langle 0|T\{S_{n}^{i}S_{\bar{n}}^{i}\}(0)\delta^{(2)}\Big(\vec{k}_{s\perp}-\vec{\mathcal{P}}_\perp^{SD}\Big)\bar{T}\{S_{n}^{i}S_{\bar{n}}^{i}\}(0)|0\rangle \rb\,.
\end{align}
where the subscript $\vec{p}_{h \perp} = 0$ indicates that the calculation of these functions is to be done in the frame in which $\vec{p}_{h \perp} = 0$.
Now in the collinear functions, the sum over states is unrestricted by the mMDT/soft-drop grooming or jet definition, but still excludes the observed hadron. All particles within the collinear function are automatically guaranteed to be within the groomed jet. The collinear-soft function $S_{i}^{\perp}$, however, contains particles which may or may not pass the grooming procedure. The operator $\vec{\mathcal{P}}^{SD}$ gives the total momentum of all particles which are included in the groomed jet, so that for a state $|X\rangle$:
\begin{align}
\vec{\mathcal{P}}^{SD}|X\rangle &=\sum_{i\in X^{SD}}\vec{p}_i|X\rangle 
\end{align}
Finally, we use the soft Wilson-line definition:
\begin{align}
S_{k}^{i}(x)&=P\text{exp}\Big(ig\int_{0}^{\infty}ds\, k\cdot A^{a}(x+s k)\mathbf{T}_{i}^{a}\Big)\,.
\end{align}
Here $i$ denotes the representation of the Wilson-line. The definition of all gauge invariant collinear operators may be found in Ref. \cite{Rothstein:2016bsq}. The substructure parameters $\mathcal{M}$ that we wish to measure in this particular case are the transverse momentum of the identified hadron, $\vec{p}_{h \perp}$, and its energy fraction, $z_h$, with respect to the ungroomed jet energy, $E_J$. So that $d\mathcal{M} = d^2\vec{p}_{h\perp} d z_h $. 
\begin{figure}[h!]
 \centerline{\includegraphics[scale = 0.4]{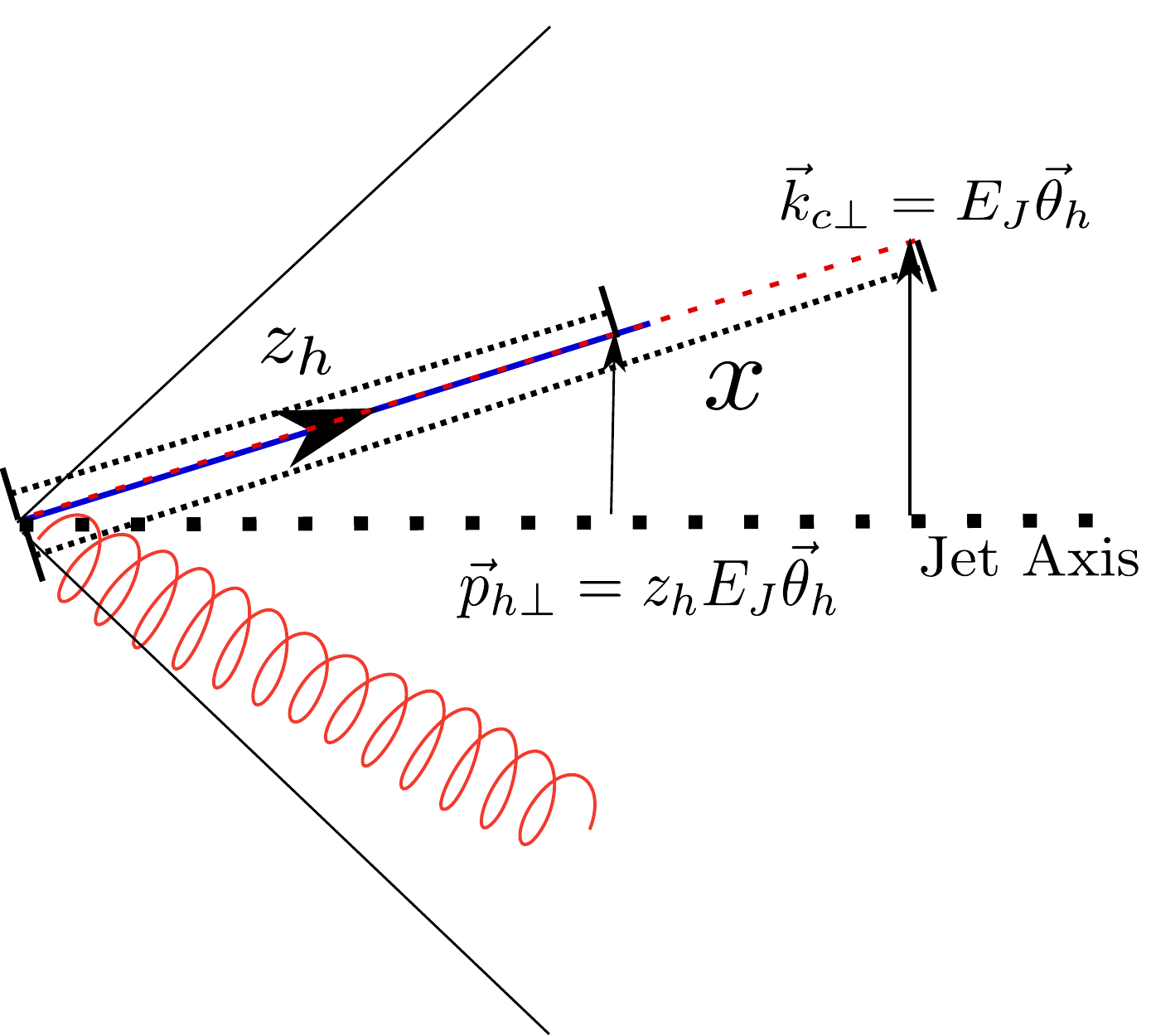}}
\caption{The geometric configuration involved in matching the TMDFF to the standard FF. $\theta_{h}$ is the angle that both the hadron (blue line) with momentum fraction $z_h$ of the jet and its initiating parton (red dotted line) make to the jet axis, and is set by perturbative splittings up to power corrections. The initiating parton has momentum fraction $x$ of the jet, so that the fragmented hadron has momentum fraction $\frac{z_h}{x}$ of the initiating parton.}
\label{fig:matching_geometry}
\end{figure}
The $p_{\perp}$ dependent collinear function (TMDFJF), $\mathcal D_{i/h}$, also implicitly depends on the hadronization scale $\Lambda_{\text{QCD}}$. For perturbative values of $p_{\perp}$, we can separate out the long distance non-perturbative physics at $\Lambda_{\text{QCD}}$ by matching the TMDFJF onto an ordinary fragmentation function. We illustrate the geometry of this matching in Figure \ref{fig:matching_geometry}:
\begin{align}
\label{matching}
\mathcal{D}_{i/h}^{\perp}\Big(z_h,\vec{k}_{c\perp},E_J\Big) = \int_{z_h}^1 \frac{dx}{x} \mathcal{J}_{ij}^{\perp}\left(x, \vec{k}_{c\perp}, E_J \right) D_{j/h}\left(\frac{z_h}{x}\right)\,.
\end{align}
The operator definition of the fragmentation function is \footnote{Keeping in mind that the light-like direction $n$ in the fragmentation function is not the same as the direction $n$ in the TMDFF, as illustrated in  Figure \ref{fig:matching_geometry}.}:
\begin{align}
D_{q/i}\Big(z_h,E_q\Big)&=  z_i\sum_{X} \frac{1}{2N_c} \delta( 2E_q - p_{X}^{-}-p_i^-)\text{tr}\left[\frac{\slashed {\bar n}}{2} \langle 0|\chi_n (0)|X i\rangle \langle X i|\bar \chi_n(0)|0\rangle \right]_{\vec{p}_{i \perp} = 0} \nn
\end{align}
with $z_i = E_i/E_q$ is the energy fraction of the final state i with respect to the fragmenting quark.
\begin{align}
D_{g/i}\Big(z_i,E_g\Big)&= z_i\sum_{X} \frac{1}{2(N_c^2-1)} \delta( 2E_g - p_{X}^{-}-p_i^-) \text{tr}\left[\langle 0|B_{n\perp}^{\mu} (0)|X i\rangle \langle X i|B_{n\perp\mu}(0)|0\rangle \right] _{\vec{p}_{i \perp} = 0} \,.
\end{align}
The final form of the factorized cross- section can now be written as:
\begin{align}
\label{factorization}
z_h^2\frac{d\sigma}{dz_h d^2p_{h\perp}}&=\sum_{i=g,q}F_i\Big(Q,R,\zcut,\vec{p}_J,C,\mu\Big) \int d^2 \vec{\tilde k}_{c \perp} \int d^2 \vec{k}_{s \perp} \delta^2 \left( \frac{\vec{p}_{h\perp}}{z_h}+ \vec{\tilde k}_{c \perp}+ \vec{k}_{s\perp}\right) \nn\\
& \times S_i^{\perp}\left( \vec{k}_{s\perp}, E_J,\zcut \right)
 \int_{z_h}^1 \frac{dx}{x} \mathcal{J}^{\perp}_{ij}\left(x, \vec{k}_{c\perp}, E_J\right)_{h.f} D_{j/h}\left(\frac{z_h}{x}\right)+...\,.
\end{align}
We can also arrive at this result by starting off with a more conventional definition of the jet function as described in Appendix \ref{sec:Alt}.
\section{Renormalization and Resummation}
Given that the modes in Eq. \eqref{eq:power_counting} of the factorization have the same invariant mass, the renormalization is to be performed within the context of SCET$_{\text{II}}$. The bare functions enjoy both ultra-violet and rapidity divergences that must be renormalized, along the lines of Ref. \cite{Chiu:2012ir} (see also \cite{Collins:2011zzd} and \cite{Becher:2010tm}). Taking the Fourier transform of all functions, thus going to the so-called $b$-space distributions, we write:
\begin{align}
\mathcal{D}_{i/h}^{\perp}\Big(z_h,\vec{b}_{\perp},E_J\Big)&=Z^{c}_{i}\Big(\frac{E_J}{\nu},\mu\vec{b}_{\perp},\alpha_s(\mu)\Big)\,\mathcal{D}_{i/h}^{\perp}\Big(z_h,\frac{E_J}{\nu},\mu\vec{b}_{\perp},\alpha_s(\mu)\Big)\,,\\
S_{i}^{\perp}\Big(\vec{b}_{\perp},E_J,\zcut\Big)&=Z^{s}_{i}\Big(\frac{\zcut E_J}{\nu},\mu\vec{b}_{\perp},\alpha_s(\mu)\Big)\,S_{i}^{\perp}\Big(\frac{\zcut E_J}{\nu},\mu\vec{b}_{\perp},\alpha_s(\mu)\Big)\,.
\end{align} 
Likewise, the quark and gluon fractions in Eq. \eqref{eq:SD_factorization} are renormalized:
\begin{align}
F_i\Big(Q,R,\zcut,\vec{p}_J,C\Big)&=Z^{F}_{i}\Big(\frac{\zcut E_J}{\mu},\alpha_s(\mu)\Big)F_i\Big(Q,R,\frac{\zcut E_J}{\mu},\vec{p}_J,C,\alpha_s(\mu)\Big).
\end{align} 
 The full factorization structure of these fractions is in general complicated and unknown (potentially suffering from both factorization violating contributions and non-global logarithms), but they do not essentially effect the predicted shape of the distribution for a quark \emph{or} a gluon. As the physical cross-section itself is renormalization group invariant, we have the general constraint on the renormalization factors:
\begin{align}
Z^{c}_{i}\Big(\frac{E_J}{\nu},\mu\vec{b}_{\perp},\alpha_s(\mu)\Big)Z^{s}_{i}\Big(\frac{\zcut E_J}{\nu},\mu\vec{b}_{\perp},\alpha_s(\mu)\Big)&=\Big(Z^{F}_{i}\Big(\frac{\zcut E_J}{\mu},\alpha_s(\mu)\Big)\Big)^{-1}\,,
\end{align}
where $Z^F_i$ is the renormalization factor for the flavor fraction $F_i$. Thus we may write:
\begin{align}
  \label{eq:SD_factorization_II}
z_h^2\frac{d\sigma}{dz_h d^2p_{h\perp}}&=\sum_{i=g,q}F_i\Big(Q,R,\zcut,\vec{p}_J,C,\mu\Big)\int db b J_0(b \vec{p}_{h\perp}/z_h)\mathcal{D}_{i/h}^{\perp}\Big(z_h,\frac{E_J}{\nu},\mu\vec{b}_{\perp},\alpha_s(\mu)\Big)\nn\\
&\times S_{i}^{\perp}\Big(\frac{\zcut E_J}{\nu},\mu\vec{b}_{\perp},\alpha_s(\mu)\Big)+...\,,
  \end{align}
  where b $= |\vec{b}_{\perp}|$ and $J_0$ is the zeroth order Bessel function of the first kind. In general these functions obey a set of renormalization group (RG) equations of the form:
\begin{align}
\mu\frac{d}{d\mu}G(\mu,\nu)&=\gamma_{\mu} G(\mu,\nu)\,,\\
\nu\frac{d}{d\nu}G(\mu,\nu)&=\gamma_{\nu} G(\mu,\nu)\,,
\end{align}
where $G$ can be either $S_{i}^{\perp}$ or $\mathcal{D}_{i/h}^{\perp}$. The first is the standard RG equation for the ultraviolet ($\mu$) anomalous dimension, whereas the second yields the rapidity ($\nu$) anomalous dimension. Only the functions describing the fragmentation process have rapidity renormalization group running, whereas the parton fractions do not. The renormalized functions possess $\mu$ anomalous dimensions of the following form:
\begin{align}
\gamma_{\mu,i}^{F}&= -\Gamma_{\text{cusp}}^i[\alpha_s(\mu)]\; \ln \zcut^2 +\gamma^F_{i}[\alpha_s(\mu)]\,,\\
\gamma_{\mu,i}^{\mathcal{D}}&= \Gamma_{\text{cusp}}^i[\alpha_s(\mu)]\; \ln \frac{\nu^2}{(2E_J)^2}+\gamma^{\mathcal{D}}_{i}[\alpha_s(\mu)]\,,\\
\gamma_{\mu,i}^{S}&= -\Gamma_{\text{cusp}}^i[\alpha_s(\mu)]\; \ln \frac{\nu^2}{\zcut^2 (2E_J)^2}+\gamma^{S}_{i}[\alpha_s(\mu)]\,.
\end{align}
We can also write down the all orders form of the $\nu$ anomalous dimensions. The hard factors $F_i$ are independent of $\nu$. The $\nu$ anomalous dimension for the collinear function ($\mathcal{D}_{i/h}^{\perp}$) is equal in magnitude but opposite in sign to that of the collinear soft function ($S_i^{\perp}$) and obeys the following consistency condition: 
\begin{align}
\frac{d}{d\ln \mu} \gamma_{\nu,i}^{S} = \frac{d}{d\ln \nu} \gamma_{\mu,i}^{S}\,.
\end{align}
This allows us to write down 
\begin{align}
\label{nuS}
 \gamma_{\nu,i}^{S}(\mu) = -2\int_{1/b_0}^{\mu} d \ln \mu' \Gamma_{\text{cusp}}^i [\alpha_s(\mu') ] + \gamma^{r}(1/b_0)\,.
\end{align}
where $b_0$ is a boundary condition in b space which is usually set to $b_0 = b e^{\gamma_E}/2$.
\subsection{All orders resummation}

To resum Large logarithms in $p_{h\perp}/E_J$, we first run the collinear soft function in $\nu$ from its natural scale $ \nu_s \sim 2E_J \zcut$ to the scale $\nu_c \sim 2E_J$. Then we run both the collinear and collinear soft function in $\mu$ from an appropriately chosen low scale $\mu_L$ to the high scale $\mu_H \sim E_J$.
Ulsing Eq.(\ref{nuS}) the evolution in $\nu$ then yields, 
\begin{align}
S^{\perp}_i(\mu,\nu= 2E_J ) = S^{\perp}_i(\mu, \nu =2E_J \zcut ) \text{Exp} \lb \ln(z^2_{cut})\lp \int_{1/b_0}^{\mu}d \ln \mu' \Gamma^i_{\text{cusp}}[\alpha_s(\mu')]+ \gamma^{r}(1/b_0)\rp \rb\,,
\end{align}
Next we want to evolve $\mathcal{D}_{i/h}^{\perp}$ and $S^{\perp}_i$ in $\mu$ from the scale $\mu_L$ to $\mu_H$. The $\mu$ anomalous dimensions for these functions to all orders are,
\begin{align}
\gamma_{\mu,i}^{\mathcal{D}} & = +2 \Gamma_{\text{cusp}}^i[\alpha_s(\mu)] \ln \lp \frac{\nu}{2E_J}\rp +\gamma^{\mathcal{D}}_i[\alpha_s(\mu)]\,,\nn\\
\gamma_{\mu,i}^{S } & = -2\Gamma_{\text{cusp}}^i[\alpha_s(\mu)]   \ln  \lp \frac{\nu}{2E_J \zcut }\rp+\gamma^{S}_i [\alpha_s(\mu)]\,.
\end{align}
The combined $\mu$ anomalous dimension for these function is given as 
\begin{align}
 \gamma_{\mu,i}^{\mathcal{D}} +\gamma_{\mu,i}^{S}&= 2 \Gamma_{\text{cusp}}^i[\alpha_s(\mu)] \ln \lp z_{cut}\rp+\gamma^{\mathcal{D}}_i[\alpha_s(\mu)]+ \gamma^{S}_i[\alpha_s(\mu)]\,\nn\\
& = 2 \Gamma_{\text{cusp}}^i[\alpha_s(\mu)] \ln \lp z_{cut}\rp-\gamma^{F}_i[\alpha_s(\mu)]
\end{align}
The evolution kernel is given as 
\begin{align}
\label{evolution}
&\mathcal{D}_{i/h}^{\perp}(\mu_H,\nu =2E_J) S^{\perp}_i(\mu_H,\nu=2E_J) =   U_i(\mu_L,\mu_H) \times \lb \mathcal{D}_{i/h}^{\perp}(\mu_L,\nu =2E_J ) S^{\perp}_i(\mu_L,\nu=2E_J \zcut ) \rb \,,
\end{align}
where the resummation exponent $U_i$ is:
\begin{equation}
  U_i(\mu_L,\mu_H) \equiv \text{Exp} \lb -\int_{\mu_L}^{\mu_H} d \ln\mu \gamma^{F}_i[\alpha_s(\mu)] +2  \ln(\zcut)  \lp \int_{1/b_0}^{\mu_H}d \ln \mu  \Gamma_{\text{cusp}}^i[\alpha_s(\mu)]+ \gamma^{r}(1/b_0)\rp \rb\,.
  \end{equation}
The point to be noted is that the term multiplying $\ln \zcut $ in the exponent is the all orders rapidity anomalous dimension.
\subsection{Resummation to NLL accuracy}
The resummation exponent we have to all orders is 
\begin{align}
  U_i&= \text{Exp} \lb -\int_{\mu_L}^{\mu_H} d \ln\mu \lp \gamma^{F}_i[\alpha_s(\mu)]\rp +2  \ln(\zcut) \int_{1/b_0}^{\mu_H}d \ln \mu  \lp \Gamma_{\text{cusp}}^i[\alpha_s(\mu)]+ \gamma^{r}(1/b_0)\rp \rb \nn\\
&= \text{Exp}\lb \int_{\mu_L}^{\mu_H} d \ln\mu \lp -\gamma^{F}_i[\alpha_s(\mu)] +2 \ln(\zcut) ( \Gamma_{\text{cusp}}^i[\alpha_s(\mu)]+ \gamma^{r}(1/b_0))\rp \rb\nn\\
& \times\text{Exp}\lb 2 \ln(\zcut) \lp \int_{1/b_0}^{\mu_L}d \ln \mu  \Gamma_{\text{cusp}}^i[\alpha_s(\mu)]+ \gamma^{r}(1/b_0)\rp \rb\,.\nn\\
\end{align}
At NLL, we have, 
\begin{align}
\gamma^{S}_i[\alpha_s(\mu)] &=0\,, & \ \ \ \ \ \gamma^{F}_i[\alpha_s(\mu)]&= -\frac{\alpha_s(\mu) C_i}{\pi} \bar \gamma_i\,.
\end{align}
with $\bar{\gamma}_i$ given in Eq.(\ref{eq:gammabars}). The cusp anomalous dimension can be written to all orders in perturbation theory as follows,
\begin{align}
\Gamma_{\text{cusp}}^i = C_i \sum_{n=1} \lp \frac{\alpha_s}{4 \pi} \rp^n\Gamma_n\;.
\end{align}
Relevant for the NLL result are $\Gamma_{0}$ and $\Gamma_1$ given by, 
\begin{align}
\Gamma_0 &= 4 \,, & \Gamma_1 &= 4C_A\left(\frac{67}{9}-\frac{\pi^2}{3}\right)-\frac{80}{9}T_R n_f\,.
\end{align}

In order to proceed we do one more approximation. Assuming that $\ln (\mu_L b_0)$ is small, we can do an expansion in the second exponent in this log keeping only the leading order term at NLL and ignoring non-perturbative contributions to the anomalous dimension.
Then at NLL we are left with:
\begin{align}
U_i(\mu_L,\mu_H) &= \text{Exp} \lb  \int_{\mu_L}^{\mu_H} d \ln\mu \lp  2\ln(\zcut)\Gamma_{\text{cusp}}^i(\alpha_s(\mu))-\gamma^{F}_i[\alpha_s(\mu)]\rp \rb \text{Exp} \lb \Gamma_{\text{cusp}}^i[\alpha_s(\mu_L)] \ln(\zcut^2)\ln(\mu_L b_0) \rb\nn\\
&= e^{K_i(\mu_L,\mu_H)}\left(\mu_L b_0\right)^{\omega_J^i(\mu_L, \zcut^2)}\,\,,
\end{align}
with $\omega_J^i = \ln(\zcut^2)\Gamma^i_{\text{cusp}}[\alpha_s(\mu_L)]$.
The function $K_i$ is evaluated including the running of $\alpha_s$ to two loops and is defined as 
\begin{align}
K_i(\mu_0,\mu) &= -C_i\ln(\zcut^2) \frac{\Gamma_0}{2 \beta_0} \Big \{ \ln r +\frac{\alpha_s(\mu_0)}{4 \pi}\left(\frac{\Gamma_1}{\Gamma_0} -\frac{\beta_1}{\beta_0}\right)\Big \}- \frac{\gamma_0}{2\beta_0} \ln r \,,
\end{align}\\
where we have defined $r = \alpha_s(\mu)/\ \alpha_s(\mu_0)$ and $ \gamma^{\mathcal{D}}_i = \alpha_s(\mu)/(4\pi) \gamma_0$. We can now go to momentum space by doing the inverse Fourier transform: 
\begin{align}
 \int d^2b e^{- i \vec{ q_T}/z_h \cdot b } = 2 \pi \int b db J_0(b q_T/z_h)\,,
\end{align}
\begin{align}
\tilde U (q_T) = - 2\pi e^{K_i(\mu_L,\mu_H)} \frac{z_h^2\omega_J^i}{q_T^2} \left(\frac{z_h \mu_L e^{\gamma_E}}{ q_T}\right)^{\omega_J^i}\frac{\Gamma\left[1+\frac{\omega_J^i}{2}\right]}{\Gamma\left[1-\frac{\omega_J^i}{2}\right]}\,.
\end{align}

In the perturbative regime, we set the scale $\mu_L =q_T/(z_h e^{\gamma_E})$ and $\mu_H \sim E_J$ so that we are left with:
\begin{align}
 \tilde U_i(q_T) = - 2\pi e^{K_i(\mu_L, E_J)} z_h^2\frac{\omega_J^i}{q_T^2}\frac{\Gamma\left[1+\frac{\omega_J^i}{2}\right]}{\Gamma\left[1-\frac{\omega_J^i}{2}\right]}\,.
\end{align}
There are no fixed order terms to be included at NLL. The cross section now looks like: 
\begin{align}
z_h^2\frac{d \sigma}{d z_{h} d^2 p_{h \perp}} = \sum_{i=g,q}F_i\Big(Q,R,\zcut,\vec{p}_J,C,\mu\Big) \tilde U_i(p_{h\perp}) D_{i/h}(z_h, \mu_L ) \,.
\end{align}

\section{Numerical results and extraction of Non-perturbative physics}

In this section we discuss how non-perturbative contributions can be incorporated in our formalism in Fourier space. We illustrate that for $k_T \gtrsim 4$ GeV non-perturbative contributions are highly suppressed and the resummed transverse momentum spectrum can be evaluated directly in the momentum space as was done in the previous section at NLL accuracy.
The cross section looks like 
\begin{align}
z_h^2\frac{d\sigma}{dz_h d^2p_{h\perp}}&=\sum_{i=g,q}F_i\Big(Q,R,\zcut,\vec{p}_J,C,\mu\Big) \int d^2 \vec{k}_{c \perp} \int d^2 \vec{k}_{s \perp} \delta^2 \left( \frac{\vec{p}_{h\perp}}{z_h}+ \vec{k}_{c \perp}+ \vec{k}_{s \perp}\right) \nn\\
& \times S_i^{\perp}\left( \vec{k}_{s \perp}, z_{cut}\right)\mathcal{D}_{i/h}^{\perp}\left(z_h, \vec{k}_{c \perp}\right)+...\,.
\end{align}
For convenience of notation we use $\vec{k}_{\perp} = \vec{p}_{h\perp}/z_h$ so that the cross section becomes 
\begin{align}
\frac{d\sigma}{dz_h d^2\vec{k}_{\perp}} = \sum_{i=g,q}F_i\Big(Q,R,\zcut,\vec{p}_J,C,\mu\Big) \cG_{i/h}(z_h, \vec{k}_{\perp} ,E_J,\zcut;\mu) \,,
\end{align}
where:\footnote{We have essentially recombined the factorized functions to reform, up to power corrections, the initial jet function in Eq. \eqref{eq:unfactorized_SD_jet_function}. In order to not introduce more functions, we keep the same symbol, simply dropping functional dependencies that are power suppressed.}
\begin{align}
 \cG_{i/h}(z_h, \vec{k}_{\perp} ,E_J,\zcut;\mu_L) = \int d^2 \vec{k}_{c \perp} \int d^2 \vec{k}_{s \perp} \delta^2 \left(\vec{k}_{\perp}+ \vec{k}_{c \perp}+ \vec{k}_{s \perp}\right) S_i^{\perp}\left( \vec{k}_{s \perp}, z_{cut}\right)\mathcal{D}_{i/h}^{\perp}\left(z_h, \vec{k}_{c \perp}\right)\,.
\end{align}

Taking the Fourier transform of $\cG(z_h,\vec{k}_{\perp},E_{J},\zcut)$ with respect to $\vec{k}_{\perp}$ we get,
\begin{align}
 \cG_{i/h}(z_h, b ,E_J,\zcut;\mu_L) &\equiv \int \frac{d \vec{k}_{\perp}}{(2 \pi)^2} e^{-i\vec{k}_{\perp} \cdot \vec{b}_{\perp}} \cG_{i/h}(z_h,\vec{k}_{\perp} ,E_J,\zcut;\mu_L) \nn\\
	&= \mathcal{D}_{i/h}^{\perp} ( z_h, b , E_J;\mu_L, \nu= 2E_J ) S^{\perp}_i(b ,E_J,\zcut;\mu_L, \nu= 2E_J),
\end{align}
where $\mu_L$ is taken to be a perturbative scale and the soft function $S^{\perp}_i$ is evolved in rapidity space from $2E_J \zcut $ to $2E_J$. This evolution is described as before using the RRG anomalous dimension, $\gamma_{\nu,i}^{S}(b,\mu)$, 
\begin{equation}
 S^{\perp}_i(b,E_J,\zcut;\mu_L, \nu=2E_J) = S^{\perp}_i (b,E_J,\zcut;\mu_L,\nu =2E_J z_{cut}) \text{Exp} \lp -\gamma_{\nu,i}^{S}(b,\mu_L) \ln z_{cut} \rp,
\end{equation}
where
\begin{equation}
 \gamma_{\nu,i}^{S}(b ,\mu) \equiv \int \frac{d^2 \vec{p}_{\perp}}{(2 \pi)^2} \text{Exp}\lb i \vec{p}_{\perp} \cdot \vec{b}_{\perp}  \rb \gamma_{\nu,i}^{S}(\vec{p}_{\perp},\mu)\;,
 \end{equation}
The all orders expression for $\gamma_{\nu,i}^{S}$ is (Eq. \ref{nuS})
\begin{align}
\gamma_{\nu,i}^{S}(\mu) = -2\int_{1/b_0}^{\mu} d \ln \mu' \Gamma_{\text{cusp}}^i [\alpha_s(\mu')] + \gamma^{r}(1/b_0)\,.
\end{align}
The choice of $b_0$ that minimizes all logarithms is $b_0 =b e^{\gamma_E}/2$. However, this choice enters the non-perturbative regime for $b\gtrsim \Lambda_{\text{QCD}}$. In the large b region we need a model function for the unknown non-perturbative physics which ultimately needs to be extracted from experiment. To that end, we separate out the perturbative contribution to the anomalous dimension from the non-perturbative one by defining a new b dependent scale
 \begin{equation} 
 \mu=\mu_b \equiv \frac{2 \exp(\gamma_E)}{b_*}\;, 
 \end{equation}
where $b_*= b/\sqrt{1+(b/\bmax)^2},$ and $\bmax$ is chosen such that $\mu_b$ is a perturbative scale for all values of $b$. At low values of b, $b^*$ is just b while at large values this approaches the fixed scale $b_{max}$. The replacement $b \to b_*$ is compensated with a non-perturbative model function, $g_K(b)$, to be determined from experimental data. That is,
\begin{align}
\gamma_{\nu,i}^{S}(\mu) = -2\int_{\mu_b}^{\mu} d \ln \mu' \Gamma_{\text{cusp}}^i[\alpha_s(\mu')] + \gamma_f^{i}(\mu) -g_K(\bl{b};\bmax) \,.
\end{align}
where $\gamma_f^{i}(\mu)$ is the perturbative non-cusp rapidity anomalous dimension of the collinear-soft function. This term only starts at two loops and hence we will set it to 0 for our analysis at NLL accuracy. What we have done is to put in all the non-perturbative parts of the anomalous dimension into the function $g_K$. Notice that this function depends on the precise choice of $b_{max}$ which decides the boundary between perturbative and non-perturbative physics. Also, in order to reproduce the perturbative result for small $b$, we need to impose $g_K(b \to 0) =0$.

Following Eq. \ref{evolution}, we can write the full resummed result as 
 \begin{equation}
\cG_{i/h}(z_h, b , E_J,\zcut;\mu) = \cG^{\;\text{FO}}_{i/h}(z_h, b , E_J;\mu_L) \text{Exp} \lb \gamma_{\nu,i}^{S}(b,\mu) \ln(\zcut) - \int_{\mu_L}^{\mu} d\ln(\mu') \gamma^{F}_i[\alpha_s(\mu')]\rb\;,
 \end{equation}
 
 To proceed, we need to make a choice for the scale $\mu_L$. Naturally we would like to make the canonical choice which minimizes the large logarithms in the perturbative expansion of $\mathcal{D}_{i/h}^{\perp}$ and $S_{i}$, but unfortunately in Fourier space this scale is $2 \exp(-\gamma_E)/b$ and the perturbative expansion fails for large $b$. For this reason we choose $\mu_L = \mu_b$ and compensate (as we did for $\gamma_{\nu}^{S_i}$) with a non-perturbative input function $g_{i/h}(z_h,b)$ .
 \begin{equation}
  \label{eq:finbspace}
 \cG_{i/h}(z_h, b , E_J ,\zcut;\mu) = \cG^{\;\text{FO}}_{i/h}(z_h, b , E_J;\mu_L=\mu_b)
  \text{Exp} \lb  \gamma_{\nu,i}^{S}(b ,\mu) \ln(\zcut) - g_{i/h}(z_h, b)  - \int_{\mu_b}^{\mu} \frac{d\mu'}{\mu'}\gamma^{F}_i[\alpha_s(\mu')] \rb,
 \end{equation}

 This is our final result for the maximum perturbative input. In contrast with the function $g_{i/h}$, which is specific to the particular fragmentation process, $g_{K}$ is universal for all TMD distributions and controls the non-perturbative evolution in rapidity. In this work we are primarily interested in $g_{K}$ and for what follows we take $g_{i/h}=0$. As an explicit example, we consider the process $e^+ +e^- \rightarrow$ dijets. We groom one of the jets and identify a pion in that jet. In this case then, we only have quark initiated jets and henceforth we will assume the parton initiating the jet to be a quark. The normalization factor $F_q(Q, R, z_{cut})$ in this case can be factorized for exclusive hemisphere jets, and has been evaluated to two-loop accuracy in Ref. \cite{Frye:2016okc}. Since gluon jets will not appear at leading power, the particular value of the normalization will not matter for the shape of the TMD-distribution. Then we can make predictions in the low $p_{h\perp}$ regime for extracting out the non-perturbative physics.

\begin{figure}[h!]
 \centerline{\includegraphics[width = \textwidth]{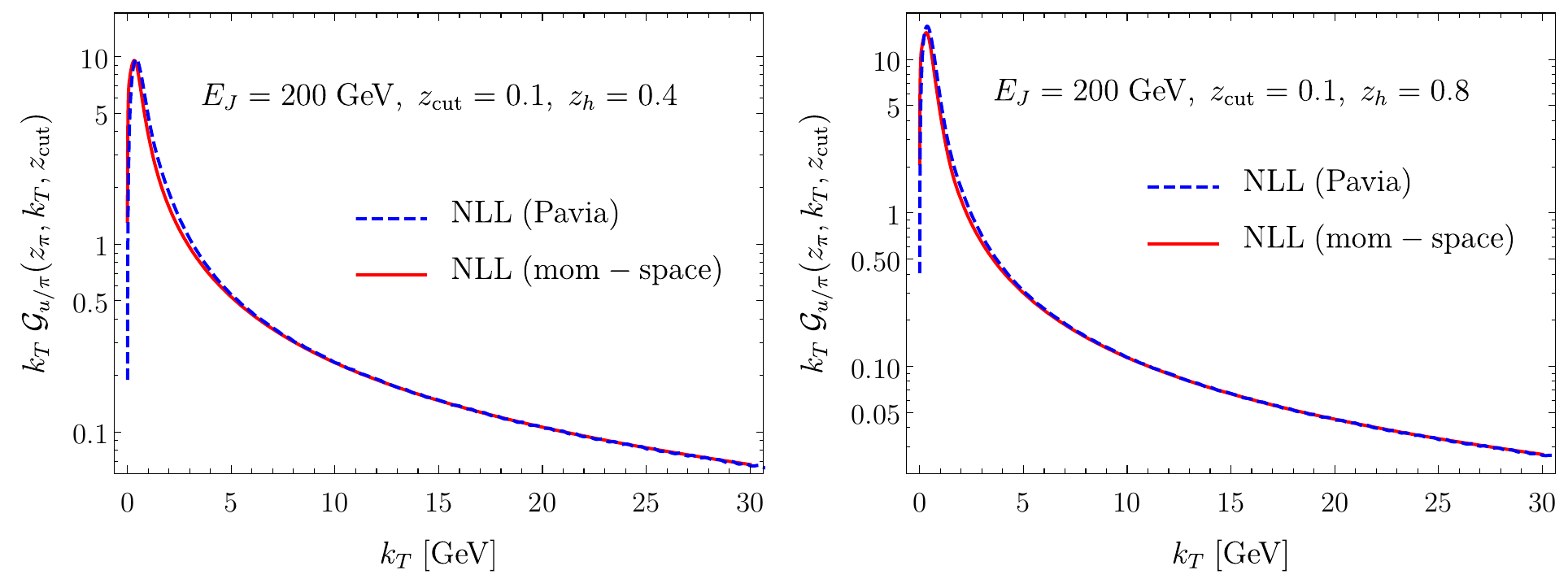}}
\caption{The up-quark to charged pions TMDFJF at NLL for $z_h=0.4$ (left) and $z_h=0.8$ (right). Other kinematic variables are described in the plots. The dashed (blue) curves corresponds to the calculation in momentum space using the Pavia fits for the CSS model. The parameters for this model are given in Table \ref{tb:models}. The solid curves correspond to the momentum space calculation for which we used a profile function to avoid the divergence of the coupling constant at small values of the factorization scale. The HKNS fragmentation functions $D_{u/\pi^+}(x;\mu)$ are taken from Ref.~\cite{Hirai:2007cx}}
\label{fig:Gg}
\end{figure}

In Figure~\ref{fig:Gg} we compare the resummed distribution evaluated in momentum space directly against our final result in Eq.(\ref{eq:finbspace}). The inverse Fourier transform was performed numerically after integrating analytically over the azimuthal angle. For the non-perturbative model function, we use the CSS model,
 \begin{equation}
  \label{eq:CSS}
 g_K( b ;\bmax) = \frac{1}{2}g_2(\bmax) b ^2 \;,
\end{equation}
 where $g_2(\bmax)$ is a free parameter to be determined by fitting to data.\footnote{Note that the implicit dependence of $g_{K}$ on the parameter $\bmax$ is absorbed into the value of $g_2$, that is to say the choice of $\bmax$ influences the fit of $g_2$. An alternative approach is to fit $g_2$ and $\bmax$ simultaneously. This approach was implemented by the Pavia 2016 fits.} The values for the parameters we use are extracted from experimental data in  Ref.~\cite{Bacchetta:2017gcc}
and are as follows: $g_2 = 0.12$ and $\bmax =1.123 $ GeV. We find that for $k_{\perp} \gtrsim 4$ GeV the two distributions are essentially identical.

\begin{figure}[h!]
 \centerline{\includegraphics[width = 0.55 \textwidth]{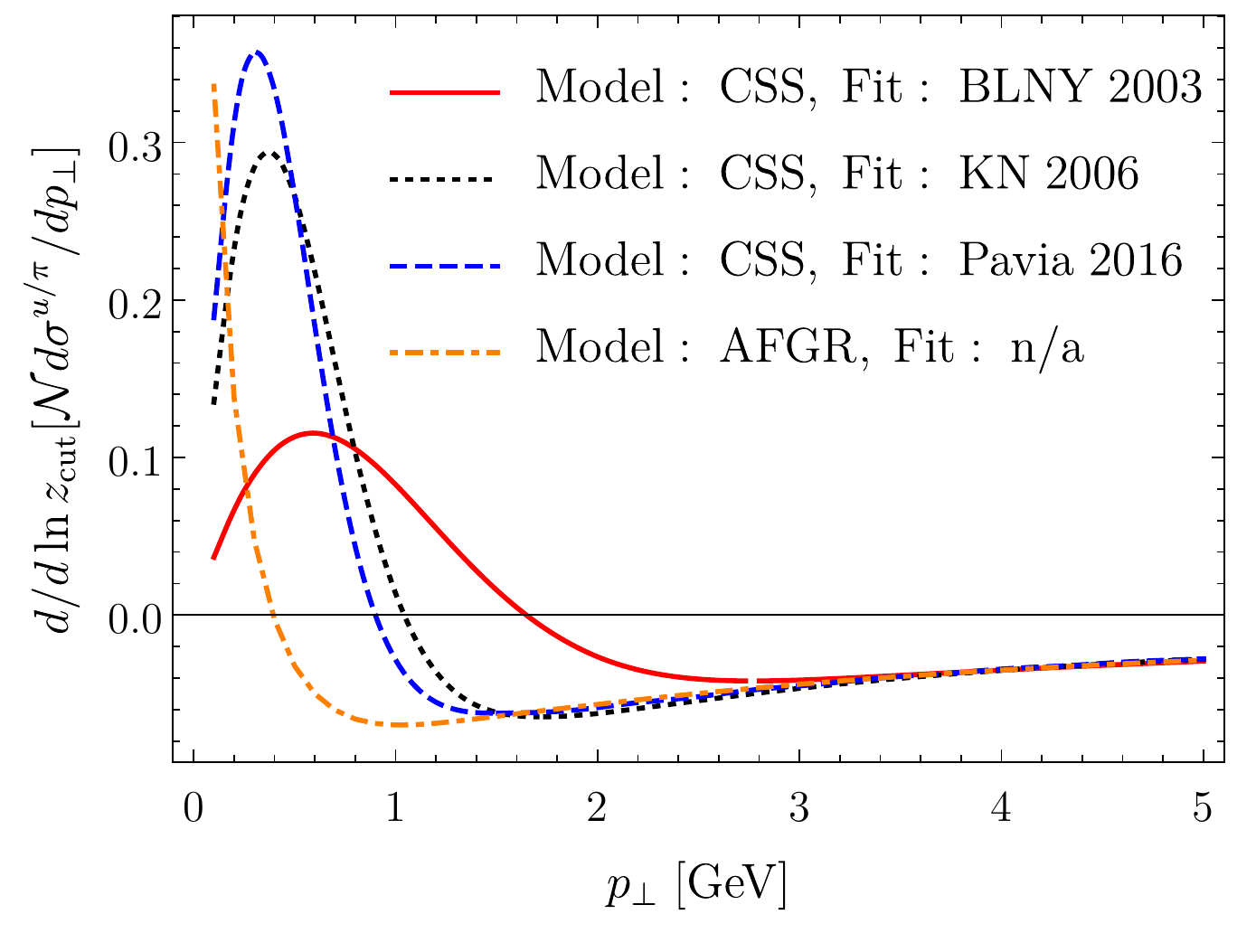}}
\caption{The logarithmic derivative of TMDFJF for the three models. All the models agree in the perturbative regime but show significant differences in the non-perturbative region.}
\label{fig:models}
\end{figure}

 We note that the only explicit dependence on $\zcut$ is in the exponent through a logarithm with a coefficient which is the rapidity anomalous dimension $\gamma_{\nu}^{S_i}(b,\mu)$. This suggests that the TMDFJF, and thus the corresponding cross section, is sensitive to the rapidity anomalous dimension through variations of $\zcut$. \footnote{For recent work on the theoretical considerations of the non-perturbative corrections to the rapidity resummation exponent, also called the collinear anomaly, see Refs. \cite{Becher:2013iya,Becher:2011xn}. One would need to also consider the effects that grooming would have on the effective rapidity range of the non-perturbative corrections. For the non-perturbative contributions including renormalon effects, see Ref. \cite{Scimemi:2016ffw}, and other field theoretic considerations see Ref. \cite{Collins:2014jpa}.}. We exploit this property through the normalized logarithmic derivative, $d/d\ln\zcut$, to discriminate between various non-perturbative models suggested previously in the literature. We evaluate the logarithmic derivative of the cross section as a function of the transverse energy for a fixed value of the energy fraction, $z_h$, for four different parametrizations of the function $g_K(b;\bmax)$. We consider three fits of the CSS model (see Eq.(\ref{eq:CSS})): 1) BNLY from Ref.~\cite{Landry:2002ix}, 2) KN from Ref.~\cite{Konychev:2005iy}, and 3) Pavia from Ref.~\cite{Bacchetta:2017gcc}. The values for the parameters of these fits are given in Table~\ref{tb:models}. We also consider the following functional form,
 \begin{equation}
  g_K(b ;\bmax) = \frac{g_2(\bmax) b_{\text{NP}}^2}{2} \ln\lp 1+\frac{ b^2}{b_{\text{NP}}^2} \rp\;,
 \end{equation}
 which was suggested in Ref.~\cite{Aidala:2014hva} and we refer to as the AFGR model. For the latter, there are no fits to the free parameters ($g_2$ and $b_{\text{NP}}$) from data. Hence, we use the approximate values suggested by the authors in the corresponding publication. Our results are illustrated in Figure~\ref{fig:models}. 
We note that for large values of the transverse momentum, the four models merge to the same distribution as expected since in that regime the perturbative anomalous dimension dominates the evolution of the cross section. In contrast, for small values of the transverse momentum the four models give clearly distinguishable distributions. These results suggest that the normalized logarithmic derivative can be used for accurate and precise extraction of the model function $g_K$ describing the non-perturbative part of the rapidity anomalous dimension of TMD observables.

\begin{table}[h!]
 \renewcommand{\arraystretch}{1.2}
 \begin{center}
\begin{tabular}{|l|c|c|c|}
\hline
Model:Fits & $g_2$ & $\bmax$ [GeV$^{-1}$]& $b_{\text{NP}}$ [GeV$^{-1}$] \\ \hline \hline
CSS:BNLY 2003~\cite{Landry:2002ix}    &  ~0.68~  & 0.5  & n.a.  \\ \hline
CSS:KN 2006~\cite{Konychev:2005iy}    &  ~0.18~  & 1.5  & n.a.  \\ \hline
CSS:Pavia 2016~\cite{Bacchetta:2017gcc} &  ~0.12~  & 1.123 & n.a.  \\ \hline
AFGR: n.a.~\cite{Aidala:2014hva}     &  ~0.10~  & 0.5  & 2.0  \\ \hline
\end{tabular}
 \label{tb:models}
\caption{Parameters for models of the non-perturbative part of the rapidity anomalous dimension.}
 \end{center} 
\end{table}

\section{Conclusion}
In this paper, we propose an observable which measures the transverse momentum of an identified hadron inside a groomed jet. We use the modified mass-drop/soft-drop grooming algorithm with an energy cut-off parameter $z_{\text{cut}}$. The radiation that recoils against the hadron is factorized into collinear and collinear- soft modes in the framework of SCET$_+$. The transverse momentum of the hadron is measured with respect to the groomed jet axis which is determined by the total momentum of the radiation in the jet that passes the soft-drop condition. For perturbative values of $p_{h \perp}$, we separate out the non-perturbative physics by matching onto a fragmentation function. 

Apart from dimensional regularization, we also need to introduce a rapidity regulator to handle divergences that arise from separating two modes( the collinear and collinear-soft) that have the same invariant mass. Consequently we have RG equation in two scales $\mu$ and $\nu$. We use RG evolution in these scales to resum large logarithms in $p_{h\perp}/E_J$. The all orders expression for the resummed result reveals that the co-efficient of $\ln z_{\text{cut}}$ in the exponent is the all orders rapidity anomalous dimension $\gamma_{\nu}$. This anomalous dimension is universal, in that the same term appears in the resummation of TMDPDF's, and more traditional TMD-fragmentation processes.

However, conventional TMD observables (those that do not include grooming) are suppressed by double logarithms in $\ln (\mu b)$ in the exponent, i.e., the exponent in b space has the LL form $Exp[- \ln^2(E_J b)]$. Since b is the conjugate parameter to $p_{h \perp}$, the b space cross-section is sensitive to non-perturbative physics only at large values of b. However, the presence of the double logarithm suppresses any non-perturbative effects since the cross-section is vanishing due to perturbative contributions alone. 

The effect of including grooming is two-fold. First, it makes the shape of the $p_{h \perp}$ distribution insensitive to non-global logarithms. Second, it removes from the resummed exponent, one power of $\ln(E_J b)$, replacing it with a $-\ln(z_{\text{cut}})$. For the typical values $z_{\text{cut}}$ that are used in experiment, this is much smaller than $\ln E_J b$ at large (b $\geq $ 2 GeV$^{-1}$) values of b. The physical interpretation of this replacement is that the grooming algorithm effectively cuts off the soft radiation at a specific energy fraction of the jet that is much larger than the transverse momentum scales that normally set the soft scaling. In standard TMD-observables this soft part can have, in principle, an arbitrarily small energy fraction, up to the kinematics requiring it to be within the jet and the on-shell conditions. This automatically means that the cross-section is much more sensitive to non-perturbative physics as compared to the corresponding ungroomed observable, and in particular, this sensitivity can be probed by comparing the groomed jet with different values of $z_{\text{cut}}$.

We take advantage of this sensitivity to test the effect of several different non-perturbative models that have been proposed in literature to modify the behavior of $\gamma_{\nu}$ at large b. We then use our formalism to give a prediction for the shape of the $p_{\perp}$ distribution of a pion. An analysis of the slope (with respect to $\ln z_{\text{cut}}$) at low values of $p_{\perp}$ is a measure of the rapidity anomalous dimension. We compare the results using different non-perturbative models as shown in Fig \ref{fig:models}. The significant differences between the predictions is indicative of the discriminating power of this observable.

An important consideration for hadron-hadron colliders (currently the Large Hadron Collider or the Relativistic Heavy Ion Collider, where there is significant interest in transverse momentum dependent observables, see Refs. \cite{Aschenauer:2013woa,Aschenauer:2015eha,Aschenauer:2016our}) is the $z_{\text{cut}}$ dependence of the quark and gluon fractions. It may be the precision of determining the quark and gluon fractions for a given process that will limit the ability to probe the non-perturbative contributions to rapidity evolution.\footnote{Though looking at jets with different underlying hard processes, like $pp\rightarrow Z+j+X$ versus single inclusive jet production, and examining different jet $p_T$ bins can help give sufficiently diverse set of fractions favoring quarks or gluons.} Unlike the $e^+e^-$ case considered in the previous sections, it cannot be simply normalized away. Finding the TMD-distributions at several $z_{\text{cut}}$ is a more or less straightforward re-clustering of the same events. Experimental collaborations at both colliders have measured TMD-observables without grooming, and in particular TMD-fragmentation has been measured at ATLAS (Ref. \cite{Aad:2011sc}) down to transverse momenta scales of the order of 1 GeV, or related observables such as the jet-shape \cite{Chatrchyan:2012mec}.\footnote{For our purposes, this is the appropriate moment with respect to $z_h$ of our TMD-fragmentation functions. } Our analysis is easily extended to next-to-next leading logarithm accuracy in the resummation of the groomed fragmenting jet function, using the results of Ref. \cite{Frye:2016aiz}. The largest unknown, then, is the quark and gluon fraction functions. They could potentially be extracted directly from experiment at a specific renormalization point using groomed jet mass measurements and theoretical calculations of Refs. \cite{Frye:2016aiz, Marzani:2017mva, Aaboud:2017qwh}, and one may also be able to provide a robust theoretical estimation of these fractions for jets with a moderate $R$, using single inclusive jet production results while resumming the jet radius logarithms and perhaps even the $z_{\text{cut}}$ dependence.

\begin{acknowledgments}
We would like to thank Wouter Waalewijn for many enlightening discussions on fragmentation in jets. This work was supported by the U.S. Department of Energy through the Office of Science, Office of Nuclear Physics under Contract DE-AC52-06NA25396 and by an Early Career Research Award, through the LANL/LDRD Program, and within the framework of the TMD Topical Collaboration.
\end{acknowledgments}

\appendix

\section{One loop results }
In this section we gather together one loop results for the matching coefficient $\mathcal{J}_{ij}$ of the Collinear function $\mathcal{D}^{\perp}_{i/h}$ on to the fragmentation function, as well as the Collinear- Soft function $S^{\perp}_i$.
\label{apA}
\subsection{Matching onto the fragmentation functions}
\label{sec:Jij}
The operator definition of the collinear function $\mathcal{D}^{\perp}_{i/h}$ is given by Eq.\ref{collinear}.
The $k_{c\perp}$ dependent collinear function $ \mathcal{D}^{\perp}_{i/h}$ is matched onto a fragmentation function $D_{i/h}$, using the relation (Eq.\ref{matching})
\begin{align}
\mathcal{D}_{i/h}^{\perp}\Big(z_h,\vec{k}_{c\perp},E_J\Big) = \int_{z_h}^1 \frac{dx}{x} \mathcal{J}_{ij}^{\perp}\left(x, \vec{k}_{c\perp}, E_J \right) D_{j/h}\left(\frac{z_h}{x}\right)\,.
\end{align}
The matching is done at the parton level, i.e., by replacing h by an  appropriate final state parton. Moreover we are working in a frame in which the final state parton has  zero  transverse momentum. The one loop results for $\mathcal{J}_{ij}$, are identical to those obtained in Ref.~\cite{Bain:2016rrv}.
\begin{align}
\mathcal{J}_{i/j}^{\perp}(z,\vec{p}_{\perp},E_J;\mu,\nu) =&\delta_{ij} \delta(1-z) \delta^{(2)}(\vec{p}_{\perp})+ \frac{\alpha_sT_{ij}}{\pi}
\lbc \lb \delta_{ij} \delta(1-z) 2 \ln \lp \frac{2E_J}{\nu}\rp+\bar P_{ji}(z)\rb \mathcal{L}_{0}(p_{\perp},\mu) \nn \\
& +c_{ij}(z)\delta^{(2)}(\vec{p}_{\perp}) \rbc \,,
\end{align}
with 
$T_{qq}=T_{qg}=C_F$, $T_{gg}=C_A$, $T_{gq}=T_F$.
The anomalous dimensions in momentum space are 
\begin{align}
\gamma_{\mu,i}^{\mathcal{D}} = \frac{\alpha_s C_i}{\pi}\lb 2 \ln \lp \frac{\nu}{2E_J}\rp+\bar \gamma_i\rb\,,\nn\\
\gamma_{\nu,i}^{\mathcal{D}} = -(8\pi) \alpha_s C_i \mathcal{L}_0(p_{\perp},\mu)\,,
\end{align}
$\mathcal{D}_{i/h}^{\perp}$ includes the anomalous dimensions of $D_{j/h}$ and $\mathcal{J}_{ij}$.
with 
\begin{align}
  \label{eq:gammabars}
\bar \gamma_q &=3/2 \; & \bar \gamma_g&= \beta_0/(2 C_A) \,,
\end{align}
In fourier(b) space, $\gamma_{\mu}$ remains unchanged while $\gamma_{\nu}$ becomes 
\begin{align}
\gamma_{\nu,i}^{\mathcal{D}} = 2\frac{\alpha_s C_i}{\pi} \ln \lp \frac{\mu b e^{\gamma_E}}{2}\rp \,.
\end{align}

\subsection{collinear-soft function, $S_i^{\perp}$}
The operator definition of the collinear soft function is given by Eq.\ref{collinear}
\begin{align}
 S^{\perp}_i= \delta^2(\vec{p}_{\perp})+\frac{2\alpha_s}{\pi} C_i \ln \lp \frac{\nu}{2E_J \zcut }\rp\mathcal{L}_0(p_{\perp}, \mu)\,,
\end{align}
which leads to the $\mu$ and $\nu$ anomalous dimensions 
\begin{align}
\gamma_{\mu,i}^{S} = - \frac{2\alpha_s}{\pi} C_i \lp \frac{\nu}{2E_J \zcut }\rp\,,\nn\\
\gamma_{\nu,i}^{S}= (8 \pi)\alpha_s C_i \mathcal{L}_0(p_{\perp},\mu) \,.
\end{align}
In Fourier space, the $\nu$ anomalous dimension changes to 
\begin{align}
\gamma_{\nu,i}^{S}= -2\frac{\alpha_s C_i }{\pi} \ln \lp \frac{\mu b e^{\gamma_E}}{2}\rp\,.
\end{align}
We then see immediately that the consistency condition for rapidity RG invariance at one loop is satisfied 
\begin{align}
\gamma_{\nu,i}^{S}+\gamma_{\nu,i}^{\mathcal{D}}=0\,.
\end{align}

\section{An Alternative Path to Factorization of Standard FF}
\label{sec:Alt}
In this section, we provide an alternative way by which we can come to the final form of the factorized cross-section given in Eq. \ref{factorization}. We start off with a more conventional definition of the jet function:
\begin{align}\label{jetv}
&\mathcal{G}_{q/h}\Big(z_h,\vec{p}_{h \perp},\zcut,R,E_J\Big) \nonumber\\
&\qquad = z_h\sum_{X\in \text{Jet}(R)} \frac{1}{2N_c} \delta( 2E_J - p_{X}^{-}-p_h^-) \delta^{(2)}(\vec{p}_{h \perp}+\vec{p}_{X^{SD} \perp})\text{tr}\left[\frac{\slashed {\bar n}}{2} \langle 0|\chi_n (0)|X h\rangle \langle X h|\bar \chi_n(0)|0\rangle \right]  \,.
\end{align}
$X$ here contains all the particles in the jet of radius $R$  excluding the hadron, $X^{SD}$ are the subset of those particles which pass the mMDT/soft-drop grooming procedure of the state $|X,h\rangle$. In this case its clear that the transverse momentum of the hadron is the recoil against the transverse momentum of all the particles in the jet which pass the grooming procedure. 
As before, we may further factorize the jet function in Eq. \eqref{jetv} as (leaving the flavor generic):
\begin{align}
\label{Jetq}
\mathcal{G}_{i/h}\Big(z_h,\vec{p}_{h\perp},\zcut,R,E_J\Big)&= z_h\int d^2 \vec{k}_{c \perp} \int d^2 \vec{k}_{s \perp} \delta^2 \left( \vec{p}_{h\perp}+ \vec{k}_{c \perp}+  \vec{k}_{s\perp}\right)\tilde{\mathcal{D}}^{\perp}_{i/h} \left(z_h, E_J, \vec{k}_{c \perp}\right)\nn\\
&\times S_i^{\perp}\left( \vec{k}_{s\perp}, z_{cut}\right)\,.
\end{align}
The operator definitions of these functions are exactly the same as those in Eq. \ref{collinear}. For perturbative values of $p_{\perp}$, we can separate out the long distance non-perturbative physics at $\Lambda_{QCD}$ by matching the TMDFJF onto an ordinary fragmentation function:
\begin{align}
\tilde{\mathcal{D}}_{i/h}^{\perp}\Big(z_h,E_J,\vec{k}_{c\perp}\Big) = \int_{z_h}^1 \frac{dx}{x} \mathcal{J}_{ij}^{\perp}\left(x, \vec{k}_{c\perp}, z_h, \vec{p}_{h\perp}\right) D_{j/h}\left(\frac{z_h}{x}\right)\,.
\end{align}

Apart from a dependence on $\vec{k}_{c\perp}$, the matching co-efficient $\mathcal{J}^{\perp}_{ij}$ also depends on $\vec{p}_{h\perp}$ and $z_h$ via the angle ($\vec{\theta}_h =\vec{p}_{h\perp}/(z_h E_J)$) that the final state hadron makes with the groomed jet axis. While it is possible to do the matching calculation directly(i.e., in the frame where the transverse momentum is measured with respect to the groomed jet axis), it is particularly convenient to do so in a frame in which the hadron has zero transverse momentum (In this frame the matching closely resembles the matching of the TMDPDF onto a PDF). Let us call this the hadron-frame (h.f). In this frame, we are guaranteed by construction that $\mathcal{J}^{\perp}_{ij}$ only depends on $\vec{k}_{c\perp}$ and x. The details of the matching in h.f are given in Section \ref{sec:Jij}. We then rotate back to the frame of our experiment via an inverse rotation by $\vec{\theta}_h$. 
\begin{align}
\tilde{\mathcal{D}}_{i/h}^{\perp}\Big(z_h,E_J,\vec{k}_{c\perp}\Big) = \int_{z_h}^1 \frac{dx}{x} \mathcal{J}^{\perp}_{ij}\left(x, \vec{k}_{c\perp}-\vec{\theta}_h E_J (1-z_h)\right)_{h.f} D_{j/h}\left(\frac{z_h}{x}\right)\,.
\end{align}
where $\mathcal{J}_{ij}(\vec{p}_{\perp})_{h.f}$ are the matching coefficients evaluated in the hadron frame. We have also used the fact that the total energy of all the collinear final state particles except the hadron is just $(1-z_h) E_J$ up to power corrections. Defining $ \vec{\tilde k}_{c\perp} =\vec{k}_{c\perp}-\vec{\theta}_h E_J (1-z_h)$, our jet function Eq. \ref{Jet} becomes

\begin{align}
\mathcal{G}_{i/h}\Big(z_h,\vec{p}_{h\perp},\zcut,R,E_J\Big)&= \int d^2 \vec{\tilde k}_{c \perp} \int d^2 \vec{k}_{s \perp} \delta^{(2)} \left( \frac{\vec{p}_{h\perp}}{z_h}+ \vec{\tilde k}_{c \perp}+  \vec{k}_{s\perp}\right) S_i^{\perp}\left( \vec{k}_{s\perp}, z_{cut}\right)\,.\nonumber\\
\end{align}

The final form of the factorized cross- section can now be written as 
\begin{align}
z_h^2\frac{d\sigma}{dz_h d^2p_{h\perp}}&=\sum_{i=g,q}F_i\Big(Q,R,\zcut,\vec{p}_J,C,\mu\Big) \int d^2 \vec{\tilde k}_{c \perp} \int d^2 \vec{k}_{s \perp} \delta^2 \left( \frac{\vec{p}_{h\perp}}{z_h}+ \vec{\tilde k}_{c \perp}+  \vec{k}_{s\perp}\right) \nn\\
& \times S_i^{\perp}\left( \vec{k}_{s\perp}, z_{cut}\right)
 \int_{z_h}^1 \frac{dx}{x} \mathcal{J}^{\perp}_{ij}\left(x, \vec{\tilde k}_{c \perp}\right)_{h.f} D_{j/h}\left(\frac{z_h}{x}\right)+...\,.
\end{align}
which is exactly the same Eq. \ref{factorization}. The key point to be noted is that we can do the matching in any reference frame. Obviously , the matching co-efficient will change depending on which frame we choose. However, at the end, as long as we rotate back to the frame in which the total groomed jet momentum has zero transverse momentum, we will always arrive at the same result.
\bibliographystyle{JHEP}

\bibliography{./TMD_FJF}

\end{document}